\documentclass{sig-alternate}
\usepackage{epsfig}
\newtheorem{theorem}{Theorem}[section]
\newtheorem{corollary}{Corollary}[section]
\newtheorem{lemma}{Lemma}[section]
\newtheorem{proposition}{Proposition}[section]
\newtheorem{definition}{Definition}[section]
\newtheorem{non-theorem}{Non-theorem}[section]

\newcommand{\thm}{\begin{theorem}}
\newcommand{\lem}{\begin{lemma}}
\newcommand{\pro}{\begin{proposition}}
\newcommand{\dfn}{\begin{definition} \rm}
\newcommand{\rem}{\begin{remark}}
\newcommand{\xam}{\begin{example}}
\newcommand{\cor}{\begin{corollary}}
\newcommand{\prf}{\begin{proof}}
\newcommand{\ethm}{\end{theorem}}
\newcommand{\elem}{\end{lemma}}
\newcommand{\epro}{\end{proposition}}
\newcommand{\edfn}{\bbox\end{definition}}
\newcommand{\erem}{\bbox\end{remark}}
\newcommand{\exam}{\bbox\end{example}}
\newcommand{\ecor}{\end{corollary}}
\newcommand{\eprf}{\end{proof}}
\newcommand{\beqn}{\begin{equation}}
\newcommand{\eeqn}{\end{equation}}
\newcommand{\bbox}{\vrule height7pt width4pt depth1pt}
\newcommand{\commentout}[1]{}
\newcommand{\M}{{\cal M}}
\newcommand{\cS}{{\cal S}}

\newcommand{\br}{\mathit{br}}

\begin{document}
\conferenceinfo{EC'06,} {June 11--15, 2006, Ann Arbor, Michigan, USA.}
\CopyrightYear{2006}
\crdata{1-59593-236-4/06/0006}

\title{Efficiency and Nash Equilibria in a Scrip System for P2P Networks}

\numberofauthors{3}

\author{
\alignauthor Eric J. Friedman\\
\affaddr{School of Operations Research and Industrial Engineering}\\
\affaddr{Cornell University}\\
\email{ejf27@cornell.edu}
\alignauthor Joseph Y. Halpern\\
\affaddr{Computer Science Dept.}\\
\affaddr{Cornell University}\\
\email{halpern@cs.cornell.edu}
\alignauthor Ian Kash\\
\affaddr{Computer Science Dept.}\\
\affaddr{Cornell University}\\
\email{kash@cs.cornell.edu}
}

\date{ }
\maketitle

\begin{abstract}
A model of providing service in a P2P network is analyzed. It is shown
that by adding a scrip system,
a mechanism that admits a reasonable Nash equilibrium that reduces
free riding can be obtained.
The effect of varying the total amount of money (scrip)
in the system on efficiency (i.e., social welfare) is analyzed, and it
is shown that by
maintaining the appropriate ratio between the total amount of money
and the number of agents, efficiency is maximized. The work has
implications for many online systems, not only P2P networks
but also a wide variety of online forums for which scrip
systems are popular, but formal analyses have been lacking.
\end{abstract}

\commentout{ A category with the (minimum) three required fields
 }



\section{Introduction}\label{sec:intro}

A common feature of many online distributed systems is that
individuals provide services for each other.
Peer-to-peer (P2P) networks (such as Kazaa \cite{Kazaa} or
BitTorrent \cite{Bittorent}) have proved
popular as mechanisms for file sharing, and applications such as
distributed computation and
file storage are on the horizon; systems such as Seti@home
\cite{seti} provide computational assistance;  systems such as
Slashdot \cite{slashdot}
provide content, evaluations, and advice forums in which people
answer each other's questions.
Having individuals provide each other with service typically
increases the social welfare: the individual utilizing the resources
of the system derives a
greater benefit from it than the cost to the individual providing
it. However, the cost of providing service can still be nontrivial.
For example, users of Kazaa and BitTorrent may be charged for
bandwidth usage; in addition, in some file-sharing systems, there is
the possibility of being sued, which can be viewed as part of the
cost.
Thus, in many systems there is a strong incentive to become a
\emph{free rider} and benefit from the system without contributing to
it.  This is not merely a theoretical problem; studies of the Gnutella
\cite{gnutella} network have shown that almost 70 percent of users
share no files and nearly 50 percent of responses are from the top 1
percent of sharing hosts~\cite{adar00}.

Having relatively few users provide most of the service creates a
point of centralization;
the disappearance of a small percentage of users can greatly impair
the functionality of the system.  Moreover, current trends seem to
be leading to the elimination of the ``altruistic'' users on which
these systems rely. These heavy users are some of the most expensive
customers ISPs have. Thus, as the amount of traffic has grown, ISPs
have begun to seek ways to reduce this traffic.  Some universities
have started charging students for excessive bandwidth usage; others
revoke network access for it \cite{cornell}.  A number of companies
have also formed whose service is to detect excessive bandwidth
usage \cite{bandwidth}.

These trends make developing a system that encourages a more equal
distribution of the work critical for the continued viability of P2P
networks and other distributed online systems.
A significant amount of research has gone into designing reputation
systems to give preferential treatment to users who are sharing
files.  Some of the P2P networks currently in use have implemented
versions of these techniques. However, these approaches tend to fall
into one of two categories: either they are ``barter-like'' or
reputational. By barter-like, we mean that each agent bases its
decisions only on information it has derived from its own
interactions. Perhaps the
best-known example of a barter-like system is BitTorrent,
where clients downloading a file try to find other clients with
parts they are missing so that they can trade, thus creating a
roughly equal amount of work. Since the barter is restricted to
users currently interested in a single file, this works well for
popular files, but tends to have problems maintaining availability
of less popular ones.  An example of a barter-like system built on
top of a more traditional file-sharing system is the credit system
used by eMule \cite{eMule}. Each user tracks his history of
interactions with other users and gives priority to those he has
downloaded from in the past. However, in a large system, the
probability that a pair of randomly-chosen users will have
interacted before is quite small, so this interaction history will
not be terribly helpful. Anagnostakis and
Greenwald~\cite{anagnostakis04} present a more sophisticated version
of this approach, but it still seems to suffer from similar
problems.

A number of attempts have been made at providing general reputation
systems  (e.g.
\cite{guha04,gupta03,eigentrust,xiong}). The basic idea is to
aggregate each user's experience into a global number for each
individual that intuitively represents the system's view of that
individual's reputation. However, these attempts tend to suffer from
practical problems because they implicitly view users as either
``good'' or ``bad'', assume that the ``good'' users will act
according to the specified protocol, and that there are relatively
few ``bad'' users.   Unfortunately, if there are easy ways to game
the system, once this information becomes widely available, rational
users are likely to make use of it. We cannot count on only a few
users being ``bad'' (in the sense of not following the prescribed
protocol). For example, Kazaa uses a measure of the ratio of the
number of uploads to the number of downloads to identify good and
bad users.  However, to avoid penalizing new users, they gave new
users an average rating.  Users discovered that they could use this
relatively good rating to free ride for a while and, once it started
to get bad, they could delete their stored information and
effectively come back as a ``new'' user, thus
circumventing the system (see \cite{anagnostakis04} for a discussion
and \cite{FrR01} for a formal analysis of this ``whitewashing'').
Thus Kazaa's reputation system is ineffective.

This is a simple case of a more general vulnerability of such
systems to \emph{sybil} attacks \cite{sybil}, where a single user
maintains multiple identities and uses them in a coordinated fashion
to get better service than he otherwise would.  Recent work has
shown that most common reputation systems are vulnerable (in the
worst case)to such attacks \cite{ChengFriedman05}; however, the
degree of this vulnerability is still unclear.  The analyses of the
practical vulnerabilities and the existence of such systems that are
immune to such attacks remains an area of active research (e.g.,
\cite{ChengFriedman05,ZhangGoel04,linkspam}).

Simple economic systems based on a scrip or money seem to avoid many
of these problems, are easy to implement and are quite popular (see,
e.g., \cite{gupta03,fileteller02,karma03}). However, they have a
different set of problems. Perhaps the most common involve
determining the amount of money in the system. Roughly speaking, if
there is too little money in the system relative to the number of
agents, then relatively few users can afford to make request.  On
the other hand, if there is too much money, then users will not feel
the need to respond to a request; they have enough money already.  A
related problem involves handling newcomers. If newcomers are each
given a positive amount of money, then the system is open to sybil
attacks.  Perhaps not surprisingly, scrip systems end up having to
deal with standard economic woes such as inflation, bubbles, and
crashes \cite{karma03}.

In this paper, we provide a formal model in which to analyze scrip
systems.  We describe a simple scrip system and show that, under
reasonable assumptions, for each fixed amount of money there is a
nontrivial Nash equilibrium involving \emph{threshold strategies},
where an agent accepts a request if he has less than $\$k$
for some threshold $k$.%
\footnote{Although we refer to our unit of scrip as the dollar,
these are not real dollars nor do we view them as convertible to
dollars.}
An interesting aspect of our analysis is that, in
equilibrium, the distribution of users with each amount of money is
the distribution that maximizes entropy (subject to the money supply
constraint).  This  allows us to compute the money supply
that maximizes \emph{efficiency} (social welfare), given the number
of agents.  It also leads to a solution for the problem of dealing
with newcomers:~we simply assume that new users come in with no
money, and adjust the price of service (which is equivalent to
adjusting the money supply) to maintain the ratio that maximizes
efficiency. While assuming that new users come in with no money will
not work in all settings, we believe the approach will be widely
applicable. In systems where the goal is to do work, new users can
acquire money by performing work.  It should also work in Kazaa-like
system where a user can come in with some resources (e.g., a private
collection of MP3s).

The rest of the paper is organized as follows.
In Section~\ref{sec:model}, we present our formal model and observe
that it can be used to understand the effect of altruists.  In
Section~\ref{sec:nonstrategic}, we examine what happens in the game
under nonstrategic play, if all agents use the same threshold
strategy. We show that, in this case, the system quickly converges
to a situation where the distribution of money is characterized by
maximum entropy. Using this analysis, we show in
Section~\ref{sec:strategic} that, under minimal assumptions, there
is a nontrivial Nash equilibrium in the game where all agents use
some threshold strategy.  Moreover, we show in
Section~\ref{sec:dynamic} that the analysis leads to an
understanding of how to choose the amount of money in the system
(or, equivalently, the cost to fulfill a request) so as to maximize
efficiency, and also shows how to handle new users.  In
Section~\ref{sec:sybils}, we discuss the extent to which our
approach can handle sybils and collusion.  We conclude in
Section~\ref{sec:conclusion}.

\section{The Model}\label{sec:model}

To begin, we  formalize providing service in a P2P network as a
non-cooperative game.
Unlike much of the modeling  in this area, our model will model the
asymmetric interactions in a file sharing system in which the
matching of players (those requesting a file with those who have
that particular file) is a key part of the system.
This is in contrast with much previous work which uses random
matching in a prisoner's dilemma.  Such models were studied in
the economics literature \cite{Kan92, Ell94} and first applied to
online reputations in \cite{FrR01};
an application to P2P is found in~\cite{Feldman04}.

This random-matching model fails to capture some salient aspects of
a number of important settings. When a request is made, there are
typically many people in the network who can potentially satisfy it
(especially in a large P2P network), but not all can. For example,
some people may not have the time or resources to satisfy the
request.  The random-matching process ignores the fact that some
people may not be able to satisfy the request.
Presumably, if the
person matched with the requester could not satisfy the match, he
would have to defect. Moreover, it does not capture the fact that
the decision as to whether to ``volunteer'' to satisfy the request
should be made before the matching process, not after.  That is, the
matching process does not capture the fact that if someone is
unwilling to satisfy the request, there will doubtless be others who
can satisfy it. Finally, the actions and payoffs in the prisoner's
dilemma game do not obviously correspond to actual choices that can
be made. For example, it is not clear what defection on the part of
the requester means. In our model we try to deal with all these
issues.

Suppose that there are $n$ agents.  At each round, an agent is
picked uniformly at random to make a request.  Each other agent is
able to satisfy this request with probability $\beta > 0 $ at all
times, independent of previous behavior. The term $\beta$ is
intended to capture the probability that an agent is busy, or does
not have the resources to fulfill the request. Assuming that $\beta$
is time-independent does not capture the intution that being an
unable to fulfill a request at time $t$ may well be correlated with
being unable to fulfill it at time $t+1$.   We believe that, in
large systems, we should be able to drop the independence
assumption, but we leave this for future work. In any case, those
agents that are able to satisfy the request must choose whether or
not to volunteer to satisfy it. If at least one agent volunteers,
the requester gets a benefit of 1 util (the job is done) and one of
volunteers is chosen at random to fulfill the request.  The agent
that fulfills the request pays a cost of $\alpha < 1$. As is
standard in the literature, we assume that agents discount future
payoffs by a factor of $\delta$ per time unit. This captures the
intuition that a util now is worth more than a util tomorrow, and
allows us to compute the total utility derived by an agent in an
infinite game. Lastly, we assume that with more players requests
come more often. Thus we assume that the time between rounds is
$1/n$.
This captures the fact that the systems we want to model are really
processing many requests in parallel, so we would expect
the number of concurrent requests to be proportional to the number of
users.%
\footnote{For large $n$, our model converges to one in which players
make requests in real time, and the time between a player's requests
are exponentially distributed
with mean 1.  In addition, the time between requests served by a
single player is also exponentially distributed.}

Let $G(n,\delta,\alpha,\beta)$ denote this game with $n$ agents, a
discount factor of $\delta$, a cost to satisfy requests of $\alpha$,
and a probability of being able to satisfy requests of $\beta$. When
the latter two parameters are not relevant, we sometimes write
$G(n,\delta)$.

We use the following notation throughout the paper:
\begin{itemize}

\item $p^t$ denotes the agent chosen in round  $t$.

\item $B_i^t \in \{ 0,1 \}$ denotes whether agent  $i$ can
satisfy the request in round  $t$.  $B_i^t = 1$ with probability
$\beta > 0 $ and $B_i^t$ is independent of $B_i^{t'}$ for all $t'
\neq t$.

\item $V_i^t \in \{ 0,1 \}$ denotes agent  $i$'s decision about
whether to volunteer in round  $t$;  1 indicates volunteering.
$V_i^t$ is determined by agent $i$'s strategy.

\item $v^t \in \{ j ~|~ V_j^tB_j^t = 1 \}$ denotes the
agent chosen to satisfy the request.  This agent is chosen uniformly
at random from those who are willing ($V_j^t = 1$) and able ($B_j^t
= 1$) to satisfy the request.

\item
$u_i^t$ denotes agent $i$'s utility in round $t$.

A \emph{standard} agent is one whose utility is determined as
discussed in the introduction; namely, the agent gets a utility of 1
for a fulfilled request and utility $-\alpha$ for fulfilling a
request. Thus, if $i$ is a standard agent, then
$$u_i^t = \left \{
\begin{array}{lll}
1 & \mbox{if } i = p_t \mbox{ and } \sum_{j \neq i} V_j^t B_j^t > 0\\
-\alpha & \mbox{if } i = v^t\\
0 & \mbox{otherwise.}\\
\end{array} \right.$$

\item $U_i = \sum_{t = 0}^\infty \delta^{t/n} u_i^t$ denotes the total
utility for agent  $i$.  It is the discounted total of agent $i$'s
utility in each round.  Note that the effective discount factor is
$\delta^{1/n}$ since an increase in $n$ leads to a shortening of the
time between rounds.

\end{itemize}

Now that we have a model of making and satisfying requests, we use
it to analyze free riding.
Take an altruist to be someone who always fulfills requests. Agent
$i$ might rationally behave altruistically if agent $i$'s utility
function has the following form, for some $\alpha'> 0$:
$$u_i^t = \left \{
\begin{array}{ll}
1 & \mbox{if } i = p_t \mbox{ and } \sum_{j \neq i} V_j^t B_j^t > 0\\
\alpha' & \mbox{if } i = v^t\\
0 & \mbox{otherwise.}\\
\end{array} \right.$$
Thus, rather than suffering a loss of utility when satisfying a
request, an agent derives positive utility from satisfying it.
Such a utility function is a reasonable representation of the
pleasure that some people get from the sense that they provide
the music that everyone is playing. For such altruistic agents,
playing the strategy that sets $V_i^t = 1$ for all $t$ is dominant.
While having a nonstandard utility function might be one reason that
a rational agent might use this strategy,
there are certainly others.  For example a naive user of filesharing
software with a good connection might well follow this
strategy.
All that matters for the
following discussion is that there are some agents that use this
strategy, for whatever reason.

As we have observed, such users seem to exist in some
large systems.
Suppose that our system has $a$ altruists.
Intuitively, if $a$ is moderately large, they
will manage to satisfy most of the requests in the system even if
other agents do no work. Thus, there is little incentive for any
other agent to volunteer, because he is already getting full
advantage of participating in the system. Based on this intuition,
it is a relatively straightforward calculation to determine
a value of $a$
that depends only on $\alpha$, $\beta$, and $\delta$, but not the number
$n$ of players in the system,
such that the dominant strategy for all standard
agents $i$ is to never volunteer to satisfy any requests (i.e.,
$V_i^t = 0$ for all $t$).

\pro There exists an $a$
that depends only on $\alpha$, $\beta$, and $\delta$
such that, in $G(n,\delta,\alpha,\beta)$ with at least $a$
altruists, not volunteering in every round is a dominant strategy for all
standard agents. \epro

\prf
Consider the strategy for a standard
player $j$ in the presence of $a$ altruists.
Even with no money, player $j$ will get a request
satisfied with probability $1 - (1 - \beta)^a$
just through the actions of these altruists.
Thus, even if $j$ is chosen to make a request in every round, the most
additional expected utility he can hope to gain by having money is
$\sum_{k=1}^\infty (1-\beta)^a \delta^k = (1 - \beta)^a / (1 - \delta)$.
If $(1 - \beta)^a / (1 - \delta) > \alpha$ or, equivalently, if
$a > \log_{1 - \beta}(\alpha ( 1 - \delta))$, never volunteering is a
dominant strategy.
\eprf

Consider the following reasonable values for our parameters: $\beta
= .01$ (so that each player can satisfy 1\% of the requests),
$\alpha = .1$ (a low but non-negligible cost), $\delta = .9999$/day
(which
corresponds to a yearly discount factor of approximately $0.95$),
and an average of 1 request per day per player. Then we only need $a
> 1145$. While this is a large number, it is small relative to the
size of a large P2P network.

Current systems all have a
pool of users behaving like our altruists.  This means that attempts
to add a reputation system on top of an existing P2P system to
influence users to cooperate \emph{will have no effect on rational users}.
To have a fair distribution of work, these systems must be
fundamentally redesigned to eliminate the pool of altruistic users.
In some sense, this is not
a problem at all.
In a system with altruists, the altruists are presumably happy, as are
the standard agents, who get almost all their requests satisfied without
having to do any work.
Indeed, current P2P network work quite well in terms of
distributing content to people.  However, as we said in the
introduction, there is some reason to believe these altruists may
not be around forever.  Thus,
it is worth looking at what can be done to make these systems work in
their absence. For
the rest of this paper we assume that all agents are standard, and
try to maximize expected utility.

We are interested in equilibria based on a scrip system.  Each time
an agent has a request satisfied he must pay the person who
satisfied it some amount. For now, we assume that the payment is
fixed; for simplicity, we take the amount to be \$1.  We denote by
$M$ the total amount of money in the system. We assume that $M > 0$
(otherwise no one will ever be able to get paid).

In principle, agents are free to adopt a very wide variety of
strategies.  They can make decisions based on the names of other
agents or use a strategy that is heavily history dependant, and mix
these strategies freely.  To aid our analysis, we would like to be
able to restrict our attention to a simpler class of strategies.  The
class of strategies we are interested in is easy to motivate.
The intuitive reason for wanting to earn money is
to cater for
the possibility that an agent will run out
before he has a chance to earn more.
On the other hand, a rational agent with plenty of mone would not want
to work,
because by the time he has
managed to spend all his money, the util will have less value than the
present cost of working.  The natural balance between these two is a
\emph{threshold strategy}.  Let $S_k$ be the strategy where an agent
volunteers whenever
he has less than $k$ dollars and not otherwise.  Note that $S_0$ is
the strategy where the agent never volunteers. While everyone
playing $S_0$ is a Nash equilibrium (nobody can do better by
volunteering if no one else is willing to), it is an uninteresting
one.  As we will show in Section~\ref{sec:strategic}, it is sufficient
to restrict our attention to this class of strategies.

We use $K_i^t$ to denote the amount of money agent  $i$ has at time
$t$. Clearly $K_i^{t+1} = K_i^t$ unless agent $i$ has a request
satisfied, in which case $K_i^{t+1} = K_i^{t+1} - 1$ or agent $i$
fulfills a request, in which case $K_i^{t+1} = K_i^{t+1} + 1$.
Formally,
$$K_i^{t+1} = \left \{ \begin{array}{lll}
K_i^t-1 & \mbox{if }i = p^t, \sum_{j \neq i} V_j^tB_j^t >0,
\mbox{ and }
K_i^t > 0\\
K_i^t+1 & \mbox{if }i = v^t \mbox{ and } K_{p^t}^t > 0\\
K_i^t & \mbox{otherwise.}
\end{array} \right.$$

The threshold strategy $S_k$ is the strategy such that
$$V_i^t = \left \{ \begin{array}{ll}
1 & \mbox{if }K_{p^t}^t > 0 \mbox{ and } K_i^t < k\\
0 & \mbox{otherwise.}
\end{array} \right.$$

\section{The Game Under Nonstrategic Play}\label{sec:nonstrategic}

Before we consider strategic play, we examine what happens in the
system if everyone just plays the same strategy $S_k$.  Our overall
goal is to show that there is some distribution over money (i.e.,
the fraction of people with each amount of money) such that the
system ``converges'' to this distribution in a sense to be made
precise shortly.

Suppose that everyone plays $S_k$.  For simplicity, assume that
everyone has at most $k$ dollars.  We can make this assumption with
essentially no loss of generality, since if someone has more than $k$
dollars, he will just
spend money until he has at most $k$ dollars.  After this point he
will never acquire more than $k$. Thus, eventually the system will
be in such a state.  If $M \geq kn$, no agent will ever be willing
to work.  Thus, for the purposes of this section we assume that $M <
kn$.

From the perspective of a single agent, in (stochastic) equilibrium,
the agent is undergoing a random walk.  However, the parameters of
this random walk depend on the random walks of the other agents and it
is quite complicated to solve directly.  Thus we consider an alternative
analysis based on the evolution of the system as a whole.

If everyone has at most $k$ dollars, then the amount of money that
an agent has is an element of $\{ 0, \ldots, k \}$.  If there are $n$
agents, then the state of the game can be
described by identifying how much money each agent has, so we can
represent it by an element of $\mathcal{S}_{k,n} = \{0, \ldots,
k\}^{\{1, \ldots, n\}}.$ Since the total amount of money is
constant, not all of these states can arise in the game.  For example
the state where each player has \$0 is impossible to reach in any game
with money in the system. Let
$m_\mathcal{S}(s) = \sum_{i \in \{ 1 \ldots n\} } s(i)$ denote the
total mount of money in the game at state $s$,
where $s(i)$ is the number of dollars that agent $i$ has in state
$s$. We want to consider only those states where the total money in
the system is $M$, namely
$$\mathcal{S}_{k,n,M} = \{s \in \mathcal{S}_{k,n} \mid m_\mathcal{S}(s)
= M \}.$$

Under the assumption that all agents use strategy $S_k$, the
evolution of the system can be treated as a Markov chain $\M_{k,n,M}$ over
the state space $\cS_{k,n,M}$.  It is possible to move from one state
to another in a single round if by choosing a particular agent to make
a request and a particular agent to satisfy it, the amounts of money
possesed by each agent become those in the second state.  Therefore
the probability of a transition from a
state $s$ to $t$ is 0 unless there exist two agents $i$ and $j$ such
that $s(i') = t(i')$ for all $i' \notin \{i,j\}$, $t(i) = s(i) +
1$, and $t(j) = s(j) - 1$. In this case the probability of
transitioning from $s$ to $t$ is the probability of $j$ being chosen
to spend a dollar
and has someone willing and able to satisfy his request ($(1 / n) (1
- (1 - \beta)^{|\{i' \mid s(i') \ne k\}| - I_j})$ multiplied
by the probability of $i$ being chosen to satisfy his request ($1 /
(|(\{i' \mid s(i') \ne k\}| - I_j)$).
$I_j$ is 0 if $j$ has $k$ dollars and 1 otherwise (it is just
a correction for the fact that $j$ cannot satisfy his own request.)

Let $\Delta^k$ denote the set of probability distributions on $\{0,
\ldots, k\}$.
We can think of an element of $\Delta^k$ as describing the fraction of
people with each amount of money.
This is a useful way of looking at the system, since we typically don't
care who has each amount of money, but just the fraction of people that
have each amount.
As before, not all elements of $\Delta^k$ are possible, given our
constraint that the
total amount of money is $M$.
Rather than thinking in terms of the total amount of money in the
system, it will prove more useful to think in terms of the average
amount of money each player has.  Of course, the total amount of money
in a system with $n$ agents is $M$ iff the average amount that each
player has is $m = M / n$.  Let $\Delta^k_m$ denote all
distributions $d \in \Delta^k$ such
that $E(d) = m$ (i.e., $\sum_{j=0}^{k} d(j) j = m$). Given
a state $s \in \cS_{k,n,M}$, let $d^s \in \Delta^k_m$ denote the
distribution of money in $s$.
Our goal is
to show that, if $n$ is large, then there is a distribution $d^* \in
\Delta^k_m$ such that, with high probability, the Markov chain
$\M_{k,n,M}$ will almost always be in a state $s$ such that $d^s$
is close to $d^*$. Thus, agents can base their decisions about what
strategy to use on the assumption that they will be in such a state.

We can in fact completely characterize the distribution $d^*$. Given
a distribution $d \in \Delta^k$, let
$$H(d) = - \sum_{\{j: d(j) \ne 0\}} d(j) \log (d(j))$$
denote the \emph{entropy} of $d$.  If $\Delta$ is a closed convex
set of distributions, then it is well known that there is a unique
distribution in $\Delta$ at which the entropy function takes its
maximum value in $\Delta$.  Since $\Delta^k_m$ is easily seen
to be a closed convex set of distributions, it follows that there is
a unique distribution in $\Delta^k_m$ that we denote
$d^*_{k,m}$ whose entropy is greater than that of all other
distributions in $\Delta^k_m$.  We now show that, for $n$
sufficiently large, the Markov chain $\M_{k,n,M}$ is almost surely
in a state $s$ such that $d^s$ is close to $d^*_{k,M/n}$.  The
statement is correct under a number of senses of ``close''.  For
definiteness, we consider the Euclidean distance.
Given $\epsilon >
0$, let $S_{k,n,m,\epsilon}$ denote the set of states $s$ in
$\cS_{k,n,mn}$ such that $\sum_{j=0}^{k} |d^s(j) - d^*_{k,m}|^2 <
\epsilon$.

Given a Markov chain $\M$ over a state space $\cS$ and $S \subseteq
\cS$, let $X_{t,s,S}$ be the random variable that denotes that $\M$
is in a state of $S$ at time $t$, when started in state $s$.

\thm \label{thm:stable} 
For all $\epsilon > 0$, all $k$, and all
$m$, there exists $n_\epsilon$ such that
for all $n > n_\epsilon$ and all states $s \in \cS_{k,n,mn}$,
there exists a time $t^*$ (which may depend on $k$, $n$, $m$, and
$\epsilon$) such that for $t >
t^*$, we have $\Pr(X_{t,s,S_{k,n,m,\epsilon}}) >
1-\epsilon$. \ethm

\prf (Sketch) Suppose that at some time $t$, $\Pr(X_{t,s,s'})$ is
uniform for all $s'$.  Then the probability of being in a set of
states is just the size of the set divided by the total number of
states. A standard technique from statistical mechanics is to show
that there is a \emph{concentration phenomenon} around the maximum
entropy distribution \cite{Jaynes}.
More precisely, using a straightforward combinatorial argument, it
can be shown that the fraction of states not in
$S_{k,n,m,\epsilon}$ is bounded by $p(n) / e^{cn}$, where $p$ is a
polynomial.  This fraction clearly goes  to 0 as $n$ gets large.
Thus,  for sufficiently large $n$,
$\Pr(X_{t,s,S_{k,n,m,\epsilon}}) > 1 - \epsilon$
if $\Pr(X_{t,s,s'})$ is uniform.

It is relatively straightforward to show that  our Markov Chain has
a \emph{limit distribution} $\pi$ over $\mathcal{S}_{k,n,mn}$, such
that for all $s,s' \in \mathcal{S}_{k,n,mn}$, $\lim_{t \rightarrow
\infty} \Pr(X_{t,s,s'}) = \pi_{s'}$.  Let $P_{ij}$ denote the probability
of transitioning from state $i$ to state $j$.
It is easily verified by an explicit computation of the transition
probabilities that $P_{ij} = P_{ji}$ for all states $i$ and $j$.
It immediatly follows from this symmetry that $\pi_s = \pi_{s'}$, so
$\pi$ is uniform. After a sufficient amount of time, the
distribution will be close enough to $\pi$, that the probabilities
are again bounded by constant, which is sufficient to complete the
theorem.
\eprf

\begin{figure}[htbp]
\centering \epsfig{file=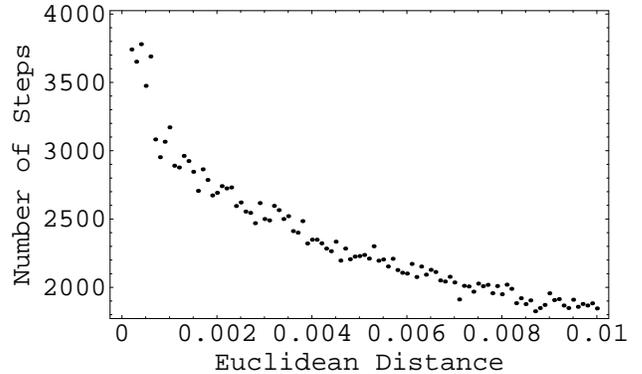, width=3.25in}
\caption{Distance from maximum-entropy distribution with 1000
agents.} \label{fig:EforN}
\end{figure}

\begin{figure}[htbp]
\centering \epsfig{file=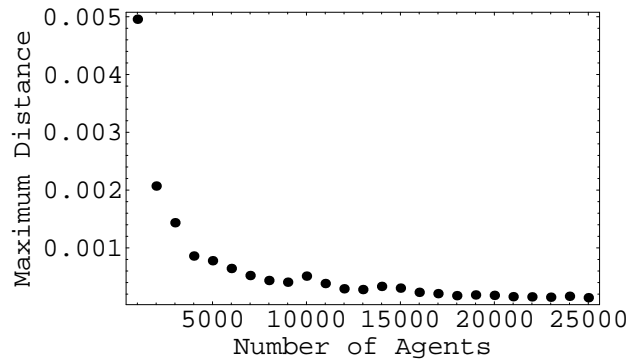, width=3.25in}
\caption{Maximum distance from maximum-entropy distribution over
$10^6$ timesteps.} \label{fig:MaxDist}
\end{figure}

\begin{figure}[htbp]
\centering \epsfig{file=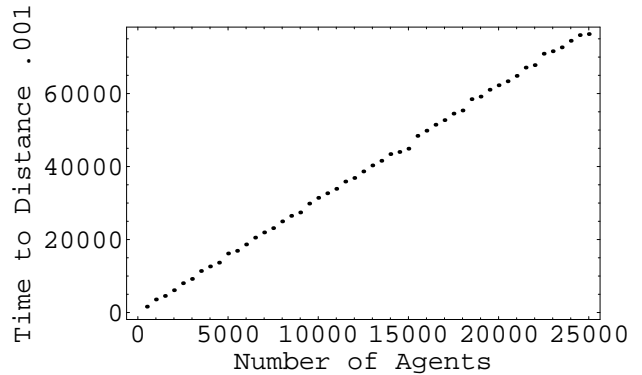, width=3.25in}
\caption{Average time to get within .001 of the maximum-entropy
distribution.} \label{fig:NforE}
\end{figure}

We performed a number of experiments that show that the maximum entropy
behavior described in Theorem~\ref{thm:stable} arises quickly for
quite practical values of $n$ and $t$.  The first experiment showed
that, even if $n=1000$, we reach the maximum-entropy distribution
quickly.
We averaged 10 runs of the Markov chain for $k = 5$ where there is
enough money for each agent to have $\$2$ starting from a very
extreme distribution (every agent has either \$0 or \$5) and
considered the average time needed to come within various distances
of the maximum entropy distribution.  As Figure~\ref{fig:EforN}
shows, after 2,000 steps, on average, the Euclidean distance from
the average distribution of money to the maximum-entropy
distribution is .008; after 3,000 steps, the distance is down to
.001.
Note that this is really only 3 real time units since with
1000 players we have 1000 transactions per time unit.

We then considered how close the distribution stays to the maximum
entropy distribution once it has reached it.
To simplify things, we
started the system in a state whose distribution was very close to
the maximum-entropy distribution and ran it for $10^6$ steps, for
various values of $n$. As Figure~\ref{fig:MaxDist} shows, the system
does not move far from the maximum-entropy distribution once it is
there.  For example, if
$n=5000$, the system is never more than distance $.001$ from the
maximum-entropy distribution; if $n=25,000$, it is never more than
$.0002$ from the maximum-entropy distribution.

Finally, we considered how more carefully how quickly the system
converges to the maximum-entropy distribution for various values of
$n$. There are approximately $k^n$ possible states, so the
convergence time could in principle be quite large.  However, we
suspect that the Markov chain that arises here is \emph{rapidly
mixing}, which
means that it will converge significantly faster (see
\cite{lovasz95} for more details about rapid mixing).  We believe
that the actually time needed is $O(n)$. This behavior is
illustrated in Figure~\ref{fig:NforE}, which shows that for our
example chain (again averaged over 10 runs), after $3n$ steps, the
Euclidean distance between the actual distribution of money in the
system  and the maximum-entropy distribution is less than .001.

\section{The Game Under Strategic Play}\label{sec:strategic}

We have seen that the system is well behaved if the agents all
follow a threshold strategy; we now want to show that there is a
nontrivial Nash equilibrium where they do so (that is, a Nash
equilibrium where all the agents use $S_k$ for some $k > 0$.)
This is not true in general.  If $\delta$ is small, then agents have
no incentive to work.  Intuitively, if future utility is
sufficiently discounted, then all that matters is the present, and
there is no point in volunteering to work.  With small $\delta$,
$S_0$ is the only equilibrium.  However, we show that for $\delta$
sufficiently large, there is another equilibrium in threshold
strategies. We do this by first showing that, if every other agent
is playing a threshold strategy, then there is a best response that
is also a threshold strategy (although not necessarily the same
one).  We then show that there must be some (mixed) threshold
strategy for which this best response is the same strategy. It
follows that this tuple of threshold strategies is a Nash
equilibrium.

As a first step, we show that, for all $k$, if everyone other than
agent $i$ is playing $S_k$, then there is a threshold strategy
$S_{k'}$  that is a best response for agent $i$.
To prove this, we need to assume that the system is close to the
steady-state distribution (i.e., the maximum-entropy distribution).
However, as long as $\delta$ is sufficiently close to 1, we can
ignore what happens during the period that the system is not in
steady state.\footnote{Formally, we need to define the strategies
when the system is far from equilibrium.  However, these far from
(stochastic) equilibrium strategies will not affect the equilibrium
behavior when $n$ is large and deviations from stochastic
equilibrium are extremely rare.}

We have thus far considered threshold strategies of the form $S_k$,
where $k$ is a natural number;  this is a discrete set of
strategies.  For a later proof, it will be helpful to have a
continuous set of strategies.
If $\gamma = k + \gamma'$, where $k$ is a natural number and $0 \le
\gamma' < 1$, let $S_\gamma$ be the strategy that performs $S_k$
with probability $1- \gamma'$ and $S_{k+1}$ with probability
$\gamma$.  (Note that we are not considering arbitrary mixed
threshold strategies here, but rather just mixing between adjacent
strategies for the sole purpose of making out strategies continuous
in a natural way.)
Theorem~\ref{thm:stable} applies to strategies $S_\gamma$ (the same
proof goes through without change),
where $\gamma$ is an arbitrary nonnegative real number.

\thm\label{thm:kprime} 
Fix a strategy $S_\gamma$ and an agent $i$.
There exists $\delta^* < 1$ and $n^*$ such
that if $\delta
> \delta^*$, $n > n^*$, and every agent other than $i$ is
playing $S_\gamma$ in game $G(n,\delta)$, then there is an integer $k'$
such that the best 
response for agent $i$ is $S_{k'}$.
Either $k'$ is unique (that is, there is a unique best response that
is also a threshold strategy), or 
there exists an integer $k''$ such
that $S_{\gamma'}$ is a best response for agent $i$ for all $\gamma'$ in
the interval $[k'',k''+1]$ (and these are the only best responses among 
threshold strategies). \ethm

\prf (Sketch:)
If
$\delta$ is sufficiently large, we can ignore what happens before
the system converges to the maximum-entropy distribution.  If $n$ is
sufficiently large, then the strategy played by one agent will not
affect the distribution of money significantly.  Thus, the
probability of $i$ moving from one state (dollar amount) to another
depends only on $i$'s strategy (since we can take the probability
that $i$ will be chosen to make a request and the probability that
$i$ will be chosen to satisfy a request to be constant).  Thus, from
$i$'s point of view, the system is a Markov decision process (MDP),
and $i$ needs to compute the optimal policy (strategy) for this MDP.
It follows from standard results \cite[Theorem 6.11.6]{Puterman}
that there is an optimal policy that is a threshold policy.

The argument that the best response is either unique or there is an
interval of best responses follows from a more careful analysis
of the value function for the MDP. \eprf

We remark that there may be best responses that are not threshold
strategies.  All that Theorem~\ref{thm:kprime} shows is that, among
best responses, there is at least one that is a threshold strategy.
Since we know that there is a best response that is a threshold
strategy, we can look for a Nash equilibrium in the space of threshold
strategies.

\thm \label{thm:br} For all $M$, there exists $\delta^* < 1$ and $n^*$
such that if
$\delta > \delta^*$ and $n > n^*$,
there exists a Nash equilibrium in the game $G(n,\delta)$ where
all agents play $S_\gamma$ for some integer $\gamma > 0$. \ethm

\prf It follows easily from the proof Theorem~\ref{thm:kprime} that if
$\br(\delta,\gamma)$ is the minimal best response threshold strategy
if all the other agents are playing $S_\gamma$ and the discount
factor is $\delta$ then, for fixed $\delta$, $\br(\delta,\cdot)$ is
a step function.
It also follows from the theorem that if there are two best
responses, then a mixture of them is also a best response.
Therefore, if we can join the ``steps'' by a vertical line, we get a
best-response curve.
It is easy to see that
everywhere that this best-response curve crosses the diagonal $y=x$
defines a Nash equilibrium where all agents are using the same
threshold strategy.  As we have already observed, one such
equilibrium occurs at 0.  If there are only \$M in the system, we
can restrict to threshold strategies $S_{k}$ where $k \le M+1$.
Since no one can have more than \$M, all strategies $S_k$ for $k >
M$ are equivalent to $S_{M}$; these are just the strategies
where the agent always volunteers
in response to request made by someone who can pay.
Clearly $\br(\delta, S_{M}) \le
M$ for all $\delta$, so the best response function is at or below
the equilibrium at $M$.
If $k \leq M/n$, every player will have at least $k$ dollars and so
will be unwilling to work and the best response is just 0.  Consider $k^*$,
the smallest $k$ such that $k > M/n$. It is not hard to show that
for $k^*$ there exists a $\delta^*$ such that for all $\delta \ge
\delta^*$, $\br(\delta,k^*) \geq k^*$.  It follows by continuity that if
$\delta \ge \delta^*$, there must be some $\gamma$ such that
$\br(\delta,\gamma) = \gamma$.  This is the desired Nash
equilibrium. \eprf

\begin{figure}[htbp]
\centering \epsfig{file=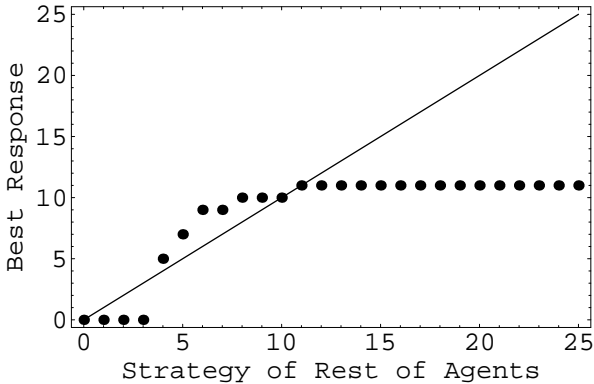, width=3.25in}
\caption{The best response function for $n=1000$ and $M=3000$.}
\label{fig:BR}
\end{figure}

This argument also shows us that we cannot in general expect
fixed points to be unique.
If $\br(\delta,k^*) = k^*$ and $\br(\delta,k+1) > k+1$
then our argument shows there must be a second fixed point.
In general there may be multiple fixed points even when $\br(\delta,k^*)
> k^*$, as illustrated in the Figure \ref{fig:BR} with $n=1000$ and
$M=3000$.

Theorem \ref{thm:br} allows us to restrict our design to agents using
threshold strategies with the confidence that there will be
a nontrivial equilibrium.  However, it does not rule out the possibility
that there
may be other equilibria that do not involve threshold stratgies.  It
is even possible (although it seems unlikely) that some of these
equilibria might be better.

\section{Social Welfare and Scalabity}\label{sec:dynamic}

Our theorems show that for each value of $M$ and $n$, for
sufficiently large $\delta$, there is a nontrivial Nash equilibrium
where all the agents use some threshold strategy $S_{\gamma(M,n)}$.
From the point of view of the system designer, not all equilibria are
equally  good; we want an equilibrium where as few as possible agents
have \$0 when they get a chance to make a request (so that they can
pay for the request) and relatively few agents have more than the
threshold amount of money (so that there are always plenty of agents
to fulfill the request).  There is a tension between these
objectives.  It is not hard to show that as the fraction of agents
with \$0 increases in the maximum entropy distribution, the fraction
of agents with the maximum amount of money decreases.  Thus, our
goal is to understand what the optimal amount of money should be in
the system, given the number of agents. That is, we want to know the
amount of money $M$ that maximizes \emph{efficiency}, i.e., the
total expected utility if all the agents use
$S_{\gamma(M,n)}$.
\footnote{If there are multiple equilibria,  we take
$S_{\gamma(M,n)}$ to be the Nash equilibrium that has highest
efficiency for fixed $M$ and $n$.}

We first observe that the most efficient equilibrium depends only on
the ratio of $M$ to $n$, not on the actual values of $M$ and $n$.

\thm\label{thm:efficiencyratio} There exists $n^*$ such that for all
games $G(n_1,\delta)$ and $G(n_2,\delta)$ where
$n_1, n_2 > n^*$, if $M_1/n_1 = M_2/n_2$, then $S_{\gamma(M_1,n_1)} =
S_{\gamma(M_2,n_2)}$. \ethm

\prf
Fix $M/n = r$.
Theorem~\ref{thm:stable}
shows that the maximum-entropy distribution depends only on $k$ and
the ratio $M/n$, not on $M$ and $n$ separately.  Thus,
given $r$, for each choice of $k$, there is a unique
maximum entropy distribution $d_{k,r}$.  The best response
$\br(\delta,k)$ depends only on the distribution $d_{k,r}$, not $M$
or $n$. Thus, the Nash equilibrium depends only on the ratio $r$.
That is, for all choices of $M$ and $n$ such that $n$ is
sufficiently large (so that Theorem~\ref{thm:stable} applies) and
$M/n = r$, the equilibrium strategies are the same. \eprf

\begin{figure}[htbp]
\centering \epsfig{file=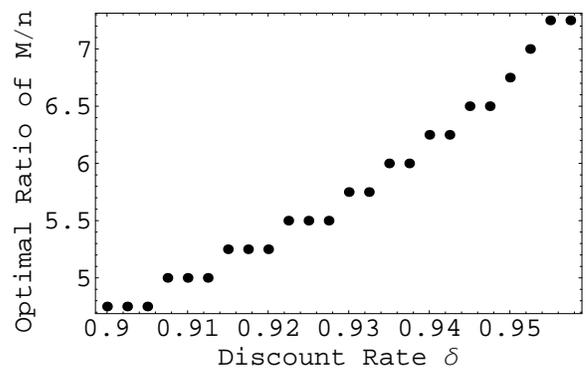, width=3.25in}
\caption{Optimal average amount of money to the nearest .25 for $\beta
= 1$}
 \label{fig:ratio}
\end{figure}

In light of Theorem~\ref{thm:efficiencyratio}, the system designer
should ensure that there is enough money $M$ in the system so that
the ratio between $M/n$ is optimal. We are currently exploring
exactly what the optimal ratio is. As our very preliminary results for
$\beta = 1$ show in Figure \ref{fig:ratio}, the ratio appears to be
monotone increasing in $\delta$, which matches the
intuition that we should provide more patient agents with the
opportunity to save more money.  Additionally, it appears to be
relatively smooth, which suggests that it may have a nice analytic
solution.

We remark that, in practice, it may be easier for the designer to
vary the price of fulfilling a request rather than injecting money
in the system.  This produces the same effect. For example, changing
the cost of fulfilling a request from \$1 to \$2 is equivalent to
halving the amount of money that each agent has.  Similarly, halving
the the cost of fulfilling a request is equivalent to doubling the
amount of money that everyone has.  With a fixed amount of money
$M$, there is an optimal product $nc$ of the number of agents and
the cost $c$ of fulfilling a request.

Theorem~\ref{thm:efficiencyratio} also tells us how to deal with a
dynamic pool of agents.
Our system can handle newcomers relatively easily: simply allow them
to join with no money.  This gives existing agents no incentive to
leave and rejoin as newcomers. We then change the price of
fulfilling a request so that the optimal ratio is maintained.
This method has the nice feature that it can be implemented in a
distributed fashion; if all nodes in the system have a good estimate
of $n$ then they can all adjust prices automatically.
(Alternatively, the number of agents in the system can be posted in a
public place.)
Approaches that
rely on adjusting the amount of money may require expensive
system-wide computations (see \cite{karma03} for an example), and must be
carefully tuned to avoid creating incentives for agents to manipulate
the system by which this is done.

Note that, in principle, the realization that the cost of fulfilling
a request can change can affect an agent's strategy.  For example,
if an agent expects the
cost to increase, then he may want to defer volunteering to fulfill
a request. However, if the number of agents in the system is always
increasing,
then the cost always decreases, so there is never any advantage in
waiting.

There may be an advantage in delaying a request,
but it is  far more costly, in terms of waiting costs than in
providing service, since we assume the need for a service is often
subject to real waiting costs, while the need to supply the service
is merely to augment a money supply. (Related issues are discussed
in \cite{FrP03}.)

We ultimately hope to modify the mechanism so that the price of a
job can be set endogenously within the system (as in real-world
economies), with agents bidding for jobs rather than
there being a fixed cost set externally. However, we have not yet
explored the changes required to implement this change.  Thus, for
now, we assume that the cost is set as a function of the number of
agents in the system (and that there is no possibility for agents to
satisfy a request for less than the ``official'' cost or for
requesters to offer to pay more than it).

\section{Sybils and Collusion}\label{sec:sybils}

In a naive sense, our system is essentially sybil-proof.
To get $d$ dollars, his sybils together still have to perform $d$
units of work.
Moreover, since newcomers enter the system with \$0, there is no
benefit to creating new agents simply to take advantage of an
initial endowment.
Nevertheless, there are some less direct ways that an agent could
take advantage of sybils.
First, by having more identities he will have a greater probability
of getting chosen to make a request.  It is easy to see that this
will lead to the agent having higher total utility.
However, this is just an artifact of our model.  To make our system
simple to analyze, we have assumed that request opportunities came
uniformly
at random.  In practice, requests are made to satisfy a desire.  Our
model implicitly assumed that all agents are equally likely to have
a desire at any particular time.  Having sybils should not increase
the need to have a request satisfied.
Indeed, it would be reasonable to assume that sybils do not make
requests at all.

Second, having sybils makes it more likely that one of the sybils
will be chosen to fulfill a request.  This can allow a user to
increase his utility by setting a
lower threshold; that is, to use a strategy $S_{k'}$ where $k'$ is
smaller than the $k$ used by the Nash equilibrium strategy.
Intuitively, the need for money is not as critical if money is
easier to obtain.
Unlike the first concern, this seems like a real issue.  It seems
reasonable to believe that when people make a decision between
a number of nodes to satisfy a request they do so at random, at least to
some extent.  Even if
they look for advertised node features to help make this decision,
sybils would allow a user to advertise a wide range of features.

Third, an agent can drive down the cost of fulfilling a request by
introducing many sybils.
Similarly, he could increase the cost (and thus the value of his money)
by making a number of sybils leave the system.  Concievably he could
alternate between these techniques to magnify the effects of work he
does.
We have not yet calculated the exact effect
of this change (it interacts with the other two effects of having
sybils that we have already noted).   Given the number of
sybils that would be needed to cause a real change in the
perceived size of a large P2P network, the practicality of this attack
depends heavily on how much sybils cost an attacker and what resources
he has available.
The second point raised regarding sybils also applies to collusion
if we allow money to be ``loaned''.  If $k$ agents collude, they can
agree that, if one runs out of money, another in the group will loan
him money.  By pooling their money in this way, the $k$ agents can
again do better by setting a higher threshold.
Note that the ``loan'' mechanism doesn't need to be built into the
system; the agents can simply use a ``fake'' transaction to
transfer the money.
These appear to be the main avenues for collusive attacks, but we
are still exploring this issue.

\section{Conclusion} \label{sec:conclusion}

We have given a formal analysis of a scrip system and have shown
that the existence of a Nash equilibrium where all agents use a
threshold strategy. Moreover, we can compute efficiency of
equilibrium strategy and optimize the price (or money supply) to
maximize efficiency. Thus, our analysis provides a formal mechanisms
for solving some important problems in implementing scrip systems.
It tells us that with a fixed population of rational users, such
systems are very unlikely to become unstable.  Thus if this stability
is common belief among the agents we would not expect
inflation, bubbles, or crashes because of agent speculation.  However,
we cannot rule out the possibility that that agents may have other
beliefs that will cause them to speculate.
Our analysis also tells us how to scale the system
to handle an influx of new users without introducing these problems:
scale the money supply to keep the average amount of money constant
(or equivalently adjust prices to achieve the same goal).

There are a number of theoretical issues that are still open,
including a characterization of the multiplicity of equilibria --
are there usually 2? In addition, we expect that one should be able
to compute analytic estimates for the best response function and
optimal pricing which would allow us to understand the relationship
between pricing and various parameters in the model.

It would also be of great interest to extend our analysis to handle more
realistic settings.  We mention a few possible extensions here:
\begin{itemize}
\item We have assumed that the world
is homogeneous in a number of ways, including request frequency,
utility, and
ability to satisfy requests.  It would be interesting to examine how
relaxing any of these assumptions would alter our results.
\item
We have assumed that there is no cost to an agent to be a member of the
system.  Suppose instead that we imposed a small cost
simply for being present in the system to reflect the costs of routing
messages and overlay maintainance.
This modification could have a significant impact on sybil attacks.
\item
We have described a scrip system that works
when there are no altruists and have shown that no system can work once
there there are sufficiently many altruists.
What happens between these extremes?
\item
One type of ``irrational'' behavior encountered with scrip systems is
hoarding.  There are some similarities between hoarding and altruistic
behavior.
While an altruist provide service for everyone, a hoarder will volunteer
for all jobs (in order to get more money) and rarely request service (so
as not to spend money).
It would be interesting to investigate the extent to which our system is
robust against hoarders.  Clearly with too many hoarders, there may not
be enough money remaining among the non-hoarders to guarantee that,
typically, a non-hoarder would have enough money to satisfy a request.
\item Finally, in P2P filesharing systems,
there are overlapping communities of various sizes that are
significantly more likely to be able to satisfy each other's
requests.
It would be interesting to investigate the effect of such communities on
the equilibrium of our system.
\end{itemize}

There are also a number of implementation issues that would have to be
resolved in a real system.  For example, we need to worry about the
possibility of agents counterfeiting money or lying about whether
service was actually provided.  Karma \cite{karma03} provdes techniques
for dealing with both of these issues and a number of others, but
some of Karma's implementation decisions point to problems for our
model.  For example, it is prohibitively expensive to ensure that
bank account balances can never go negative, a fact that our model
does not capture.
Another example is that Karma
has nodes serve as
bookkeepers for other nodes account balances.  Like maintaining a
presence in the network, this imposes a cost on the node, but unlike
that, responsibility it can be easily shirked.  Karma suggests several
ways to incentivize nodes to perform these duties.
We have not investigated whether these mechanisms be incorporated
without disturbing our equilibrium.

\section{Acknowledgements}
We would like to thank Emin Gun Sirer, Shane Henderson, Jon
Kleinberg, and 3 anonymous referees for helpful suggestions. EF, IK
and JH are supported in part by NSF under grant ITR-0325453.  JH is
also supported in part by NSF under grants CTC-0208535 and
IIS-0534064, by ONR under grant N00014-01-10-511, by the DoD
Multidisciplinary University Research Initiative (MURI) program
administered by the ONR under grants N00014-01-1-0795 and
N00014-04-1-0725, and by AFOSR under grant F49620-02-1-0101.

\bibliographystyle{abbrv}

\end{document}